\documentclass{llncs}
\usepackage[dvips]{graphics}
\usepackage{epsfig}
\usepackage{float}
\usepackage{makeidx}  
\begin{document}

\mainmatter              
\title{Deterministic aspects of Nonlinear Modulation Instability}
\author{E van Groesen, Andonowati \& N. Karjanto}
\authorrunning{Van Groesen, Andonowati and Karjanto}   

\institute{Applied Analysis \& Mathematical Physics, University of Twente,\\
POBox 217, 7500 AE, Netherlands\\
\& Jurusan Matematika, Institut Teknologi Bandung,\\
Jl. Ganesha 10, Bandung Indonesia\\
\email{groesen@math.utwente.nl, aantrav@attglobal.net, n.karjanto@math.utwente.nl}
}

\maketitle

\begin{abstract}
Different from statistical considerations on stochastic wave fields, this
paper aims to contribute to the understanding of (some of) the underlying
physical phenomena that may give rise to the occurrence of extreme, rogue,
waves. To that end a specific deterministic wavefield is investigated that
develops extreme waves from a uniform background. For this explicitly
described nonlinear extension of the Benjamin-Feir instability, the soliton on
finite background of the NLS equation, the global down-stream evolving
distortions, the time signal of the extreme waves, and the local evolution
near the extreme position are investigated. As part of the search for
conditions to obtain extreme waves, we show that the extreme wave has a
specific optimization property for the physical energy, and comment on the
possible validity for more realistic situations. \newline\textbf{Keywords}:
rogue waves, modulational instability, deterministic extreme waves,
constrained minimal energy principle.

\end{abstract}

\section{Introduction}

In this contribution we describe various aspects related to modulational,
Benjamin-Feir, instability that have been found in a detailed study of a
family of wavefields. These aspects give rise to some understanding of
phenomena that are observable in the downstream evolution of unidirectional
waves, showing the spatial evolution from a slightly modulated uniform wave
train to a position where large amplitude amplification occurs and extreme
waves arise. Although these findings are for a special class of solutions in a
simplified model, it is expected that at least some of the phenomena and
underlying physics could be quite characteristic for more general situations
in which extreme (`rogue') waves are observed. This is to be expected, since
the family of wavefields studied are the deterministic description of the
fully nonlinear evolution of the initially linear Benjamin-Feir instability
with one pair of unstable sidebands. This family is known in the NLS model as
the Soliton on Finite Background (SFB), given in \cite{AkhAnk}. Some of the
results reported in detail in Andonowati e.a., \cite{AKG}, will be put in a
broader perspective and, where possible, a link with research on stochastic
elements will be made. In fact, different from the statistical approach that
envisages the occurrence of extreme waves as a very rare occasion for which it
is not yet known if there are special circumstances that give rise to their
occurrence, we start from the opposite direction: what are the basic
underlying physical properties of extreme waves (as they appear in this
family), and extract information from this that may be characteristic for more
realistic cases too.

Slowly varying evolutions of a monochromatic wave with wavenumber and
frequency $\left(  k_{0},\omega_{0}\right)  $\ satisfying the dispersion
relation, are described in first order of the (small) wave elevations with a
complex valued amplitude $A$\ like
\[
\eta\left(  x,t\right)  \approx A\exp\left[  i\left(  k_{0}x-\omega
_{0}t\right)  \right]  +cc
\]
To study the spatial evolution, it is practical to describe the amplitude in
variables with delayed time, with the delay determined by the corresponding
group velocity $V_{0}$, so $\xi=x,\tau=t-x/V_{0}$, where $\left(  x,t\right)
$\ are the (scaled) physical laboratory variables (we suppress the first order
and second order scaling coefficients in $\tau$ and $\xi$ respectively, just
as we do in the amplitude). To incorporate dispersive and nonlinear effects in
comparable order, $A$ should satisfy in the lowest nontrivial approximation
the NLS equation, given by
\[
\partial_{\xi}A=-i\left[  \beta\partial_{\tau}^{2}A+\gamma|A|^{2}A\right]
\]
where $\beta,\gamma$ are constants, depending on the monochromatic wave. For
sufficiently large wavenumbers, both parameters have the same sign (positive
say, without restriction since the sign of $\xi$ can be changed). Then
dispersive effects (broadening of linear wavegroups) can be balanced by
focussing effects from nonlinearity; in this paper we consider only this case,
the `deep water limit'.

In section 2 we present the basic observations of the family of
SFB\footnote{Actually, in this paper we consider only the family with one pair
of initial side bands: SFB$(1)$; higher order families with $n$ pairs of
sidebands also exist, SFB$(n)$. The case $n=2$ shows phenomena like the
interaction of two SFB$(1)$ solutions, somewhat similar to interaction in NLS
of two confined soliton wave groups .
\par
{}}. We first present a description of the complete spatial evolution, using
the envelope of the waves to illustrate the development of wavegroups, and the
maximal temporal amplitude to depict the largest possible surface elevations
at each point. We investigate the time signal, and its spectral properties, at
the extreme position where the largest waves appear and describe how, as a
consequence of the appearance of phase singularities, waves in one wavegroup
are distinguished between extreme and intermittent waves that have opposite
phase. Near the extreme position, the motion of the extreme waves is studied,
showing the familiar nonlinear modification of the dispersion relation in the
physical solution, and changes in the quadratic energy spectrum in second and
higher order only. In section 3 we investigate what can be said about the
maximal possible amplitude for signals obeying certain constraints. In
particular we consider as constraints the simplest motion invariants of energy
and momentum (constants during the down stream evolution), and show that the
extreme signal has as remarkable property that it is a solution of a specific
optimization principle. Remarks and conclusions about the relevance of the
results obtained in this paper for more realistic situations will finish the paper.

\section{Nonlinear modulation instability in the SFB family}

This section is based to a considerable extent on the results presented in
\cite{AKG}; see this paper also for additional references. After some
preliminaries, a global description of downstream running nonlinearly
distorted waves according to SFB is presented, the extremal signal is studied,
and the detailed dynamics near the extreme position is investigated.

\subsection{Preliminaries}

The explicit expression for the solution of the NLS equation called Soliton on
Finite Background is given in \cite{AkhAnk}; we use the notation of
\cite{AKG}. The solution is given for the complex amplitude $A(\xi,\tau)$:%
\[
A\left(  \xi,\tau\right)  =r_{0}e^{-i\gamma r_{0}^{2}\xi}\left(  \frac
{\tilde{\nu}^{2}\cosh\left(  \sigma\xi\right)  -i\tilde{\nu}\sqrt{2-\tilde
{\nu}^{2}}\sinh(\sigma\xi)}{\cosh\left(  \sigma\xi\right)  -\sqrt{1-\tilde
{\nu}^{2}/2}\cos\left(  \nu\tau\right)  }-1\right)  .
\]
This describes actually a family of SFB solutions which depend on two
essential parameters, $r_{0}$ and $\nu$; two other parameters are related to a
shift in time and position: we will choose these such that the extreme wave
will appear for normalized variables at $x=0$, with maximal height at $t=0$.

The parameter $r_{0}$ denotes (half of) the amplitude of the uniform wavetrain
at infinity, while $\tilde{\nu}$ is a normalization of the modulation
frequency $\nu$\ of the given carrier frequency. In fact, with the notation
from \cite{AKG}, we have $\tilde{\nu}=\sqrt{\beta/\gamma}\nu/r_{0}$. Compared
to the definition of Benjamin-Feir Index $BFI$\ in \cite{Janssen03}, adapted
for the case considered here, we have the relation $\tilde{\nu}=\sqrt{2}/BFI$
so that Benjamin-Feir instability takes place for $\tilde{\nu}<\sqrt{2}$,
corresponding to $BFI>1$. The parameter $\sigma=\gamma r_{0}^{2}\sqrt
{2-\tilde{\nu}^{2}}$ happens to be the Benjamin-Feir growth factor of linear
instability theory.

In the following we will use the notation SFB to denote the solution in
physical variables, describing the surface elevation $\eta(x,t)$\ of the
physical waves without the second order Stokes effect. These second order
effects can be added; they will contribute to the actual wave heights, and
show modulations (with double modulation frequency) on the MTA described
below, but will not essentially contribute to the basic phenomenon. The role
of the second order effects in generating four wave interaction and resonance
phenomenon is already accounted for (in the considered order of accuracy) by
the NLS equation. Another important consequence of this is that in the
following we deal mainly with the wave amplitudes which directly determine the
waveheight as twice the amplitude.

In the following we will consider the spatial NLS-equation, in which case the
SFB will be periodic in time, and soliton-like in the spatial direction,
describing the spatial evolution of downstream running time-modulated waves.

For the deterministic SFB wavefield the significant waveheight $H_{s}$, a
quantity that is fundamental in the statistical description of wave fields
cannot well be defined. Yet, since the space asymptotic of the wave field is
a uniform wavetrain, when considering the averaged amplitude of the one-third
highest waves, the only consistent similar quantity would be the value
$2r_{0}$. Adopting then the (seemingly arbitrary, but often used) definition
of `rogue' wave as waves of wave height larger then $2.2\ast H_{s} $, this
will give a rough idea in which cases in the following we are dealing with
extreme waves.

\subsection{Characteristic spatial evolution}%

\begin{figure}
\begin{center}
\includegraphics[
height=1.75in,
width=4.5in
]{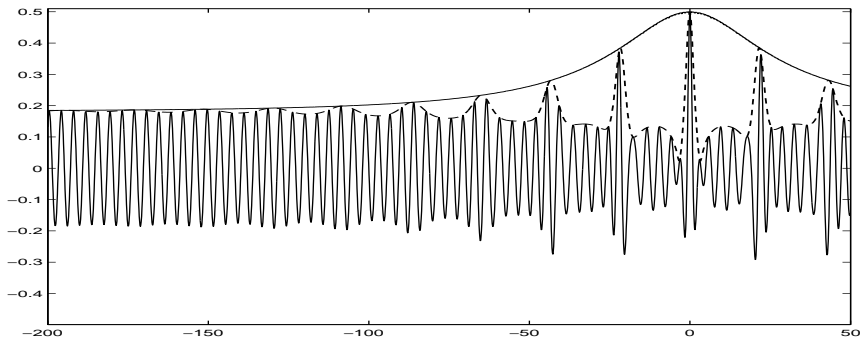}
\caption{Snapshot of the spatial wavefield SFB for $\tilde{\nu}=\sqrt{1/2}$.
Shown are the individual waves, their envelope at that instant, and the
time-independent MTA, maximal temporal amplitude. Physical dimensions are
given along the axis: horizontally the distance and vertically the surface
elevation in meters, for waves on a layer with depth of 5 meter.}%
\end{center}
\end{figure}

Fig. 1 shows a plot of a snapshot of the spatial wavefield of waves running
from left to right. At the left the slightly modulated uniform wave train
(amplitude $2r_{0}$) is seen. This modulation clearly determines a
characteristic modulation length that is maintained during the complete
downstream evolution. While moving to the right, the modulations are
amplified, creating distinct wave groups. At a certain position, called the
\textit{extreme position} (in scaled variables taken to be at $x=0$), the
largest wave appears, after which the reverse process sets in the decay
towards the asymptotic harmonic wave train (with some phase change). Note that
near the extreme position, the extreme wave is locally surrounded by waves of
much smaller amplitude, as if the total energy in one wavegroup is conserved
but with the energy redistributed between waves. In time, both the waves and
the envelope shifts to the right at different speed (the phase and group
velocity respectively). Also shown in the plot is the so-called MTA, the
\textit{maximal temporal amplitude}: this is the (time-independent) curve
determined by the maximal wave height at each point, the steady envelope of
the wavegroups. Among other things, the MTA shows the global amplification
factor, the ratio of the maximal and the asymptotic amplitude; this ratio is
maximal 3, depending on $\tilde{\nu}$, but the local amplification factor near
the extreme position can actually be much larger.

As is visible in Fig.1, the waves as they are running downstream, undergo
increasingly large oscillations in amplitude. At a fixed position the maximal
amplitude is given by the value of MTA, which is reached once every time
period $T=2\pi/\nu$. In between these successive maxima, the amplitude may be
monotone, or (for sufficiently small values of $\tilde{\nu}$), non-monotone as
is shown near the extreme position.

\subsection{The extreme signal}

The \textit{extreme signal}, the time signal at the extreme position, has
various special properties; we will denote the envelope by $S\left(
\tau\right)  $. First, this extreme signal is real, and its envelope is
strictly positive for $\tilde{\nu}>\tilde{\nu}_{crit}$, while for $\tilde{\nu
}<\tilde{\nu}_{crit}$ the envelope changes sign; here $\tilde{\nu}%
_{crit}=\sqrt{3/2}$. At times when the envelope vanishes, the phase
experiences a $\pi$-jump, causing phase singularities. In any case we observe
that at the extreme position all modes that make up the time signal are
strongly phase correlated: either all having the same phase or some having
opposite phase. More particularly, it was shown in \cite{AKG} that the
envelope $S$\ satisfies a Newton-type of equation and allows a simple
phase-plane representation. Explicitly, the equation reads%
\begin{equation}
\beta\partial_{\tau}^{2}S+\gamma S^{3}=\kappa S+\lambda\label{ESignal}%
\end{equation}
where $\kappa,\lambda$ are positive constants (depending on $\tilde{\nu}$ and
$r_{0}$). We will show in the next section that this is related to an
optimization property.%

\begin{figure}
\begin{center}
\includegraphics[
height=1.75in,
width=4.5in
]{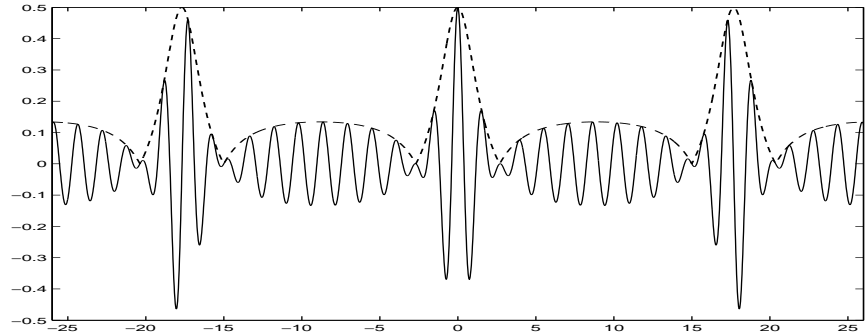}%
\caption{The extreme signal, i.e. the time signal at the extreme position
where the largest waves appear. For the same SFB parameters and scaling as in
Fig.1 the horizontal axis is the time in seconds. }%
\end{center}
\end{figure}

For a characteristic value of $\tilde{\nu}<\tilde{\nu}_{crit}$, the time
signal is plotted in Fig.2. We see that in one modulation period the wavegroup
has been split in extreme waves and a number of intermittent waves of much
smaller amplitude. The separation at times of a phase singularity, causes the
intermittent and extreme waves to have opposite phase. In the spatial plot
this shows itself in wave annihilation and wave creation at the successive
singularities (see \cite{AKG} for more details).

At the critical value $\tilde{\nu}_{crit}=\sqrt{3/2}$ \ for which the envelope
vanishes at one point and is positive at the other times in the modulation
period, the global amplification factor is precisely $2$. To satisfy the
`rogue wave' definition of amplification larger than $2.2.$ the value of
$\tilde{\nu}$ \ has to be smaller, i.e. will always correspond to the case
when phase singularities are present; for instance, for $\tilde{\nu}=1$, the
amplification is $1+\sqrt{2}\approx2.4$.

\subsection{Spectral properties of the extreme signal}

The spectral description of the time signal at a fixed position is for the
asymptotic modulated wavetrain according to Benjamin-Feir instability: the
major part of the energy in the central frequency and small contributions in
one pair of sidebands. As is to be expected form the change of the spatial
envelope, an increase of the number of relevant sidebands and large energy
exchange between the modes takes place while approaching the extremal
position; depending on the value of $\tilde{\nu}$\ the energy of the central
frequency may have been transferred to neighbouring sidebands, for $\tilde
{\nu}=\sqrt{1/2}$ even completely.

Actually, the appearance in the extreme signal of the phase singularities, and
the corresponding partitioning of the waves within one modulation period in
extreme and intermittent waves, causes that large differences in the spectrum
are observed while practically the same envelope for the extreme waves is
obtained. The intermittent waves `modulate' the spectral properties of the
extreme waves. This can be seen by writing the envelope $S(t)$\ in one period
$[0,T_{mod}]$\ as the sum of an envelope $f(t)$ of the extreme
waves, and an envelope $g(t-T_{mod}/2)$\ of the intermittent
waves centered at $T_{mod}/2=\pi/\nu$. Then the spectral
Fourier components of the complete envelope
$S(t)=f(t)-g(t-T_{mod}/2)$, the minus-sign to indicate the
$\pi$-phase difference between the waves, are given by $S_{m}=\hat{f}%
_{m}-\left(  -1\right)  ^{m}\hat{g}_{m}$, with $\hat{f}_{m},\hat{g}_{m}$ the
spectral components of $f$ and $g$ in the $m$-th sideband respectively. The
factor $\left(  -1\right)  ^{m}=\exp(im\nu T_{mod}/2)$ is a
consequence of the timeshift and has the modulational effect of decreasing and
increasing the contributions in successive sidebands, starting with a decrease
of the energy at the center frequency. In case the intermittent waves are such
that $\Sigma\left\vert S_{m}\right\vert ^{2}<\Sigma\left\vert \hat{f}%
_{m}\right\vert ^{2}$ this indicates that the presence of the intermittent
waves makes it possible that the same maximal amplitude can be obtained for
less energy. In Fig. 3 the spectra of the extreme signal are shown for three
values of $\tilde{\nu}$.%

\begin{figure}
[ptb]
\begin{center}
\includegraphics[
height=1.413in,
width=1.4006in
]{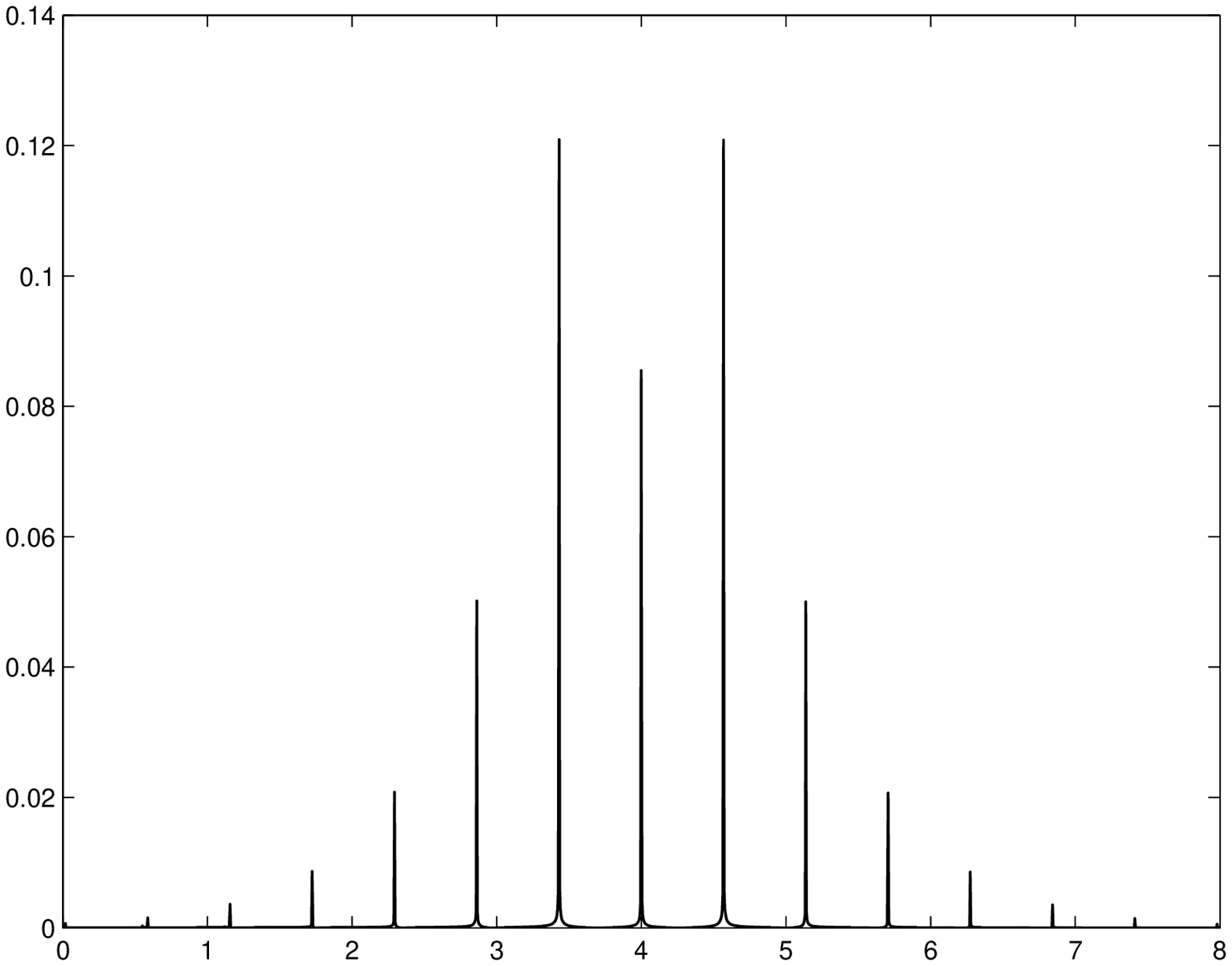}\,\,
\includegraphics[
height=1.413in,
width=1.4006in
]{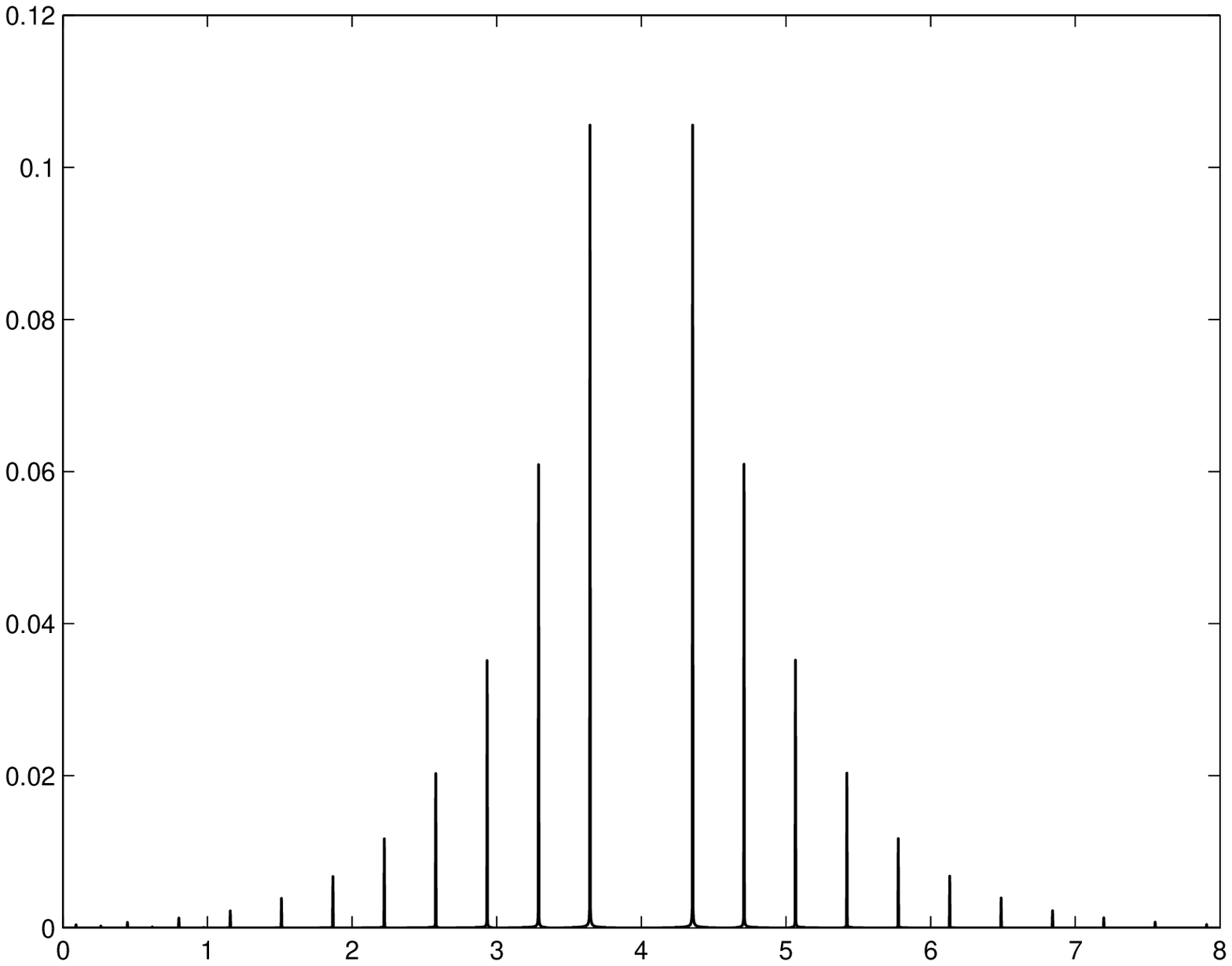}\,\,
\includegraphics[
height=1.413in,
width=1.4006in
]{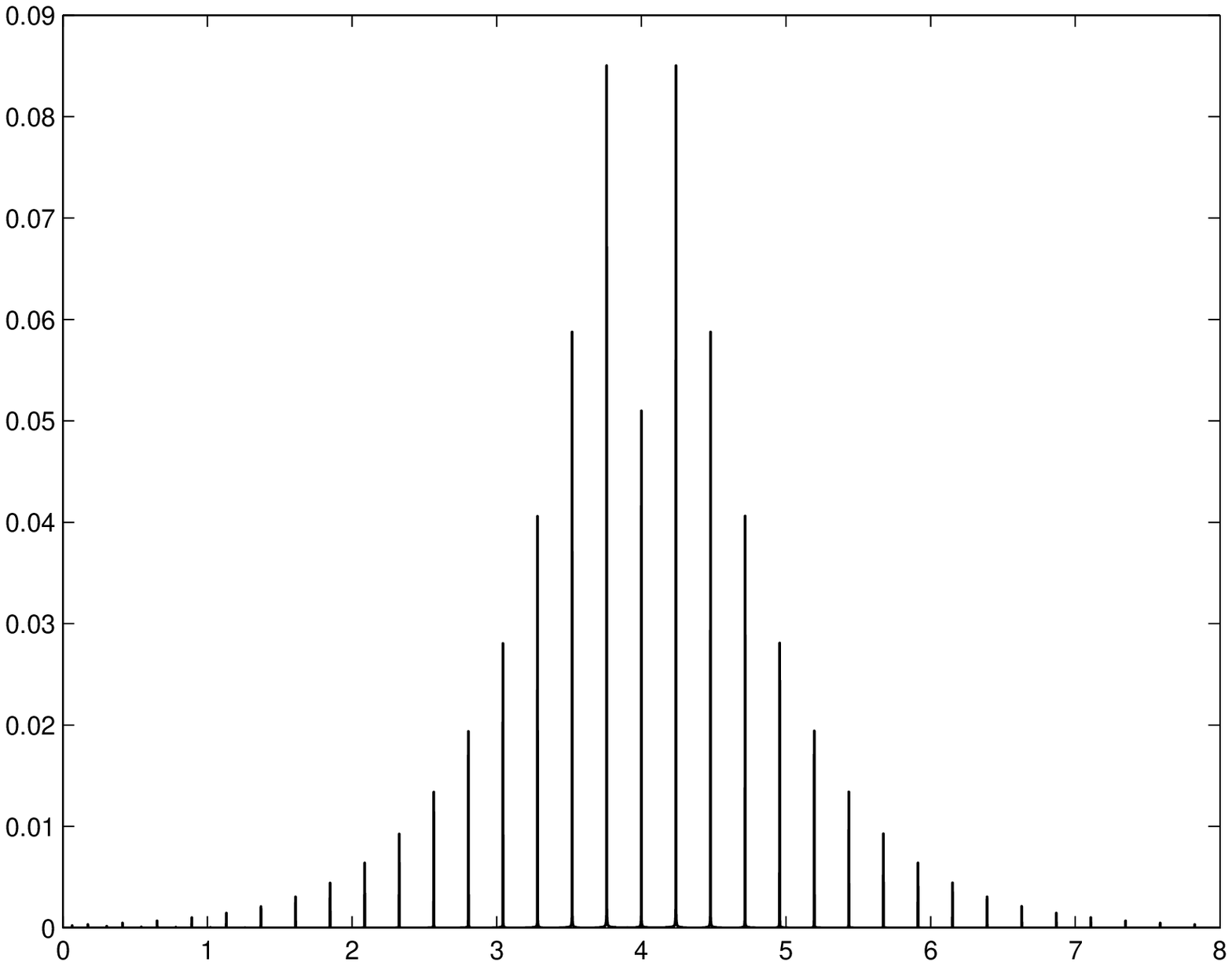}%
\caption{The absolute value of the amplitude spectrum of the extreme signal
for three values of $\tilde{\nu}=1,\sqrt{1/2},1/2$ from left to right. Observe
the vanishing contribution at the central frequency $\omega_{0}=4$ for
$\tilde{\nu}=\sqrt{1/2}$.}%
\end{center}
\end{figure}

The observation described here may be a warning when interpreting spectra for
extreme waves, and just as well when looking for conditions on spectra to
describe extreme waves. When background waves are present (the intermittent
waves in the extreme signal may be considered like that), they will greatly
disturb the spectrum related to the extreme waves and can increase the maximal
possible amplitude at given energy.

\subsection{Local evolution near the extreme position}

We investigate the evolution of the signal and its spectral components in
space near the extreme position. We show that the first order change in phase
is described by a nonlinear modification of the dispersion relation with an
additional quadratic term and a term of Fornberg-Whitham type. The change of
the envelope, and of the quadratic spectrum, is shown to be of higher order.

The change of the amplitude near the extreme position is found from a direct
Taylor expansion $A\left(  \xi\right)  =A\left(  0\right)  +\xi\left[
\partial_{\xi}A\right]  _{\xi=0}+\frac{\xi^{2}}{2}\left[  \partial_{\xi}%
^{2}A\right]  _{\xi=0}+...$. Using the evolution equation and the equation for
$S$ one gets
\[
A\left(  \xi\right)  =S+i\xi\left(  -\kappa S-\lambda\right)  -\frac{\xi^{2}%
}{2}\left[  2\gamma\kappa S^{3}+3\lambda\gamma S^{2}+\kappa^{2}S+\lambda
\kappa\right]  +O\left(  \xi^{3}\right)  .
\]
In the plane of the complex amplitude $A$, the evolution in time at $\xi
=0$\ is on the real axis, crossing the origin if there is a phase singularity.
For small $\xi$ the solution above can be described as $A\left(  \xi\right)
=e^{-i\kappa\xi}S-i\xi\lambda+O\left(  \xi^{2}\right)  $ which has the
geometric interpretation of two successive actions: the solution at $\xi=0$
that lies on the real axis is rotated around the origin over an angle
$-\kappa\xi$ and then followed by a shift along the imaginary axis over a
distance $-\lambda\xi$.

For the absolute value of the NLS solution we find up to third order
\[
\left\vert A\left(  \xi\right)  \right\vert ^{2}=S^{2}-\xi^{2}\left[
2\gamma\kappa S^{4}+3\lambda\gamma S^{3}-\kappa\lambda S-\lambda^{2}\right]
+O\left(  \xi^{3}\right)  ,
\]
while, writing $A=\left\vert A\right\vert e^{i\phi}$,\ the first order change
of the phase $\phi$ at $\xi=0$ is
\[
\partial_{\xi}\phi=-\kappa-\lambda/S+O\left(  \xi\right)  ,
\]
indicating once again the singular behaviour at the times of phase singularity
where $S$ vanishes. For the phase of the physical solution, $\psi=\phi+\left(
k_{0}x-\omega_{0}t\right)  $, this leads to $\partial_{x}\psi=k_{0}-\left(
\kappa+\lambda/S\right)  .$ Invoking the governing equation for $S$, the
result for the local wavenumber can be written like
\[
k(x)=k_{0}-\left(  \gamma S^{2}+\frac{\beta\partial_{t}^{2}S}{S}\right)
+O\left(  x\right)  .
\]
This can be interpreted as a nonlinear modification of the linear dispersion
relation with a quadratic contribution and a homogeneous term of
Fornberg-Withham type.

For the Fourier transformation with respect to time, we denote the spectral
components of $A(\xi)$ by $A_{m}\left(  \xi\right)  $ according to $A\left(
\xi,\tau\right)  =\Sigma A_{m}\left(  \xi\right)  e^{-im\nu\tau}$. Using the
fact that $S$ is real we find%

\[
A_{m}\left(  \xi\right)  =S_{m}+i\xi\left(  -\kappa S_{m}-\hat{\lambda
}\right)  -\frac{\xi^{2}}{2}\left[  2\gamma\kappa\left(  \widehat{S^{3}%
}\right)  _{m}+3\gamma\lambda\left(  \widehat{S^{2}}\right)  _{m}+\kappa
^{2}S_{m}+\kappa\widehat{\lambda}\right]  +O\left(  \xi^{3}\right)  .
\]
Here $\hat{\lambda}=\lambda\delta\left(  m\right)  $ is a contribution to the
central frequency $m=0$\ only, while $\widehat{S^{2}}$ and $\widehat{S^{3}}$
denote convolution of second and third order respectively. For the absolute
value we find up to third order
\[
\left\vert A_{m}\left(  \xi\right)  \right\vert ^{2}=S_{m}^{2}-\xi^{2}\left[
2\gamma\kappa S_{m}\cdot\left(  \widehat{S^{3}}\right)  _{m}+3\lambda\gamma
S_{m}\cdot\left(  \widehat{S^{2}}\right)  _{m}-\kappa\hat{\lambda}S_{0}%
-\hat{\lambda}^{2}\right]  +O\left(  \xi^{3}\right)  .
\]
To relate this with the change of the physical quadratic spectrum, we get for
the quadratic spectrum using $\eta\left(  x,t\right)  =\Sigma A_{m}\left(
x\right)  \exp\left[  -i\left(  \omega_{0}+m\nu\right)  t\right]  +cc:$%
\[
P_{m}\left(  x\right)  =P_{m}\left(  0\right)  -x^{2}\left[  2\gamma\kappa
S_{m}\cdot\left(  \widehat{S^{3}}\right)  _{m}+3\lambda\gamma S_{m}%
\cdot\left(  \widehat{S^{2}}\right)  _{m}-\kappa\hat{\lambda}S_{0}%
-\hat{\lambda}^{2}\right]  +O\left(  x^{3}\right)  .
\]
Writing $A_{m}=\left\vert A_{m}\right\vert e^{i\theta_{m}\left(  \xi\right)
}$ we find at $\xi=0$\ for the phase change of the spectral components of the
amplitude $\partial_{\xi}\theta_{m}\left(  0\right)  =-\kappa-\hat{\lambda
}/S_{0}$. For the phase of the physical solution $\eta$ this modifies the
change due to linear dispersion:%
\[
\partial_{x}\psi_{m}=-\kappa-\frac{\hat{\lambda}}{S_{0}}+k_{0}+m\nu/V_{0},
\]
which is the spectral version of the nonlinearly modified dispersion relation.

\section{Extremal formulations}

This section addresses some aspects centered around the question when a
wavefield (or envelope) attains its maximal value, depending on the
constraints that are imposed. We will denote this symbolically for a signal
$s\left(  t\right)  $ like%

\[
\max_{s}\,\left.  \left\{  \mathcal{A}\left(  s\right)  \, \right\vert \textnormal{\ constraints\ }\right\}  \textnormal{ with }\mathcal{A}\left(
s\right)  =\max_{t}\,s\left(  t\right)
\]
In particular, the effect of prescribing the quadratic spectrum is considered
first. Then we consider the case of relevance for the evolution of waves, when
we take as constraints motion invariants (integral quantities), and describe
that the extreme signal of the SFB solutions of the previous section arises as
special signal when the constraints are optimally chosen.

\subsection{Constrained maximal signal amplitudes}

Here we consider real functions of given period $T$ and $\nu=2\pi/T$, or look
at signals with continuous spectrum (using the notation of the latter). Any
signal with given quadratic spectrum $P\left(  \omega\right)  $\ is of the
form%
\[
s\left(  t\right)  =\int\sqrt{P\left(  \omega\right)  }e^{i\theta\left(
\omega\right)  }e^{-i\omega t}d\omega
\]
for some phase function $\theta\left(  \omega\right)  $. A completely focussed
signal would have all phases the same, say zero, $s_{foc}\left(  t\right)
=\int\sqrt{P\left(  \omega\right)  }e^{-i\omega t}d\omega,$ and produces the
signal that for the given quadratic spectrum has the largest amplitude (at
$t=0$), and%
\[
\max_{s}\,\left.  \left\{  \mathcal{A}\left(  s\right)  \,\right\vert \,s\textnormal{ has given quadratic spectrum }P\left(
\omega\right)  \, \right\}  =s_{foc}\left(  0\right)  =\int
\sqrt{P\left(  \omega\right)  }d\omega.
\]
If we relax the constraints, the results critically depend on the constraints.
For instance, if not the spectrum, but only the value of the integrated
quadratic spectrum is prescribed, the related maximization problem has no
finite solution. In Fourier language this is related to the fact that the
energy is equally partitioned over all sidebands: equipartition of energy. We
indicated in the previous section that intermittent waves can partly
contribute to a better equipartition. If stronger norms that the integrated
quadratic spectrum are prescribed, finite solutions will exist.

Most relevant seems to consider the maximization problem with constraints that
are motivated by physics. To that end we introduce the following functionals
that are related respectively to the approximation of the physical energy $H$,
the physical momentum $I$ and the `mass' functional $M$ defined by%

\[
H\left(  s\right)  =\int_{0}^{T}\left[  \frac{\beta}{2}\left(  \partial
_{t}s\right)  ^{2}-\frac{\gamma}{4}s^{4}\right]  dt,\,\,I\left(
s\right)  =\int_{0}^{T}\frac{1}{2}s^{2}dt,\,\,M\left(  s\right)
=\int_{0}^{T}s\,dt,
\]
and we investigate the optimization problem\footnote{A very interesting
statistically motivated variant of this maximization problem (without the
linear mass-constraint) has been considered by Fedele \cite{Fedele-manu04}. He
considers the initial value problem (evolution in time) and uses the
Hamiltonian and quadratic invariant functionals that are related to the
Zakharov equation, and takes as values of the constraints the values of a
linearized wavefield at an initial time.}%
\begin{equation}
\max_{s}\,\left.  \left\{  \mathcal{A}\left(  s\right)  \,\right\vert \, H(s)=h;\,I(s)=g;\,M(s)=m\,\right\}
.\label{maxA_HGM}%
\end{equation}
The resulting equation for (\ref{maxA_HGM}) follows with Lagrange multiplier
rule:%
\begin{equation}
\sigma\delta\mathcal{A}=\lambda_{1}\delta H+\lambda_{2}\delta I+\lambda
_{3}\delta M\label{EqnAHGM}%
\end{equation}
where we write the variational derivative of a functional $K$\ like $\delta K$
(when equated to zero, $\delta K=0$, this is precisely the Euler-Lagrange
equation of the functional $K$). The multipliers are related to the values of
the constraints; when the rhs doesn't vanish, the multiplier $\sigma$ can be
taken equal to one without restriction. Explicitly, the equation reads%
\[
\sigma\delta_{Dirac}\left(  t-t_{\max}(s)\right)  =\lambda_{1}\left[
-\beta\partial_{t}^{2}s+\gamma s^{3}\right]  +\lambda_{2}s+\lambda_{3}%
\]
where $t_{\max}(s)$\ is the time at which $s$ attains its maximum, and
$\delta_{Dirac}$ denotes Dirac's delta function. When the rhs vanishes, which
is consistent with $\sigma=0$, we recover the equation for the extremal signal
of SFB described above. This is only the case if the constraint values $g,h,m$
\ are chosen correctly; we show in the next subsection that this holds for the
extremal signal.

For non-optimal constraint values, the multipliers will be different and
$\sigma\neq0$. Then formally the extremal signal can still be found explicitly
but will not be a realistic physical signal since it contains a discontinuity
in the derivative at the time of maximal amplitude as a consequence of the
delta-function; the optimal solution is then obtained by pasting continuously
together parts of a suitable extreme time signal. Although the signal itself
may be non-physical, its value of the maximal amplitude provides an upperbound
for any other signal satisfying the constraints.

\subsection{Extremal property of the extreme signal}

As stated above, among signals that satisfy constrained values of the
functionals $H,I,M$ the ones with maximal amplitude will be obtained when the
constraints are such that these three functionals are linearly related (when
$\sigma=0$). We will now show that the extremal signals have this property.

Indeed, for suitable parameters $\nu,r_{0}$ related to the constraint values
$g,m$, the extreme signal will be a solution of the constrained optimization
problem for the physical energy
\begin{equation}
\min_{S}\,\left\{  \,H(S)\,\left\vert \,I(S)=g,\,M(S)=m\,\right\}  \right.  \label{minH_GM_real}%
\end{equation}

Indeed, an extremizer of this optimization problem satisfies the Lagrange
multiplier formulation
\begin{equation}
\delta H(S)=-\kappa\delta I(S)-\lambda\label{ESeqn_abstr}%
\end{equation}
for some multipliers $-\kappa,-\lambda$. Written in full for the specific
functionals, this is precisely the Newton equation (\ref{ESignal})\ for the
extreme signal. Actually, using the fact that $H$ and $I$ are restrictions to
real-valued amplitudes of motion invariants for the complex amplitude of NLS,
these values in (\ref{minH_GM_real}) are immediately found from the asymptotic
values at the uniform wavetrain; we will describe this in more detail in a
forth coming paper \cite{GroAnd05-VarExtWaves}.

\section{Conclusions and remarks}

We have studied a special class of wavefields and extracted several properties
that may be useful for the understanding of the appearance of extreme waves in
more realistic situations.

Although the wavefields are derived from the simple NLS-model and no rigorous
mathematical proof of their validity can be given, at least it can be said
that these wavefields seem to be realizable in practice, as was shown by
actual experiments performed in large wavetanks at MARIN, see the contribution
of Huijsmans e.a. \cite{HKKA04} in these proceedings. The experimental
verification is vital, in particular because most simplified models (and
certainly NLS as has been used here) do not predict breaking phenomena. When
breaking occurs, the deterministic predictions become useless, while if no
breaking occurs it can be expected that the predictions have qualitative, and
as has been shown in the experiments, also quite good quantitative, validity.

It is also appropriate to discuss the validity, the `robustness', of the basic
phenomena described here for extension to more realistic situations. In
particular this concerns the optimization property of the extreme signal: can
we expect such a property to hold in more realistic situations, without
relying on any conviction that also in those cases " ...... la nature agit
selon quelque principe d'un maximum ou minimum." (Euler, 1746).

The optimization principle involves three functionals. Two of them have a
clear physical meaning, although for NLS the complexified versions have to be
considered. These are $H$, which is when complexified the Hamiltonian of NLS
and is an approximation of the energy, and the quadratic functional $I$\ which
in complex form is also an NLS invariant that can be interpreted for the real
wavefields as the momentum. These functionals will also be present as
invariants for reliable models that are more accurate than the NLS model,
since the energy and momentum expressions are (approximations of) motion
invariants for the full surface wave equations under the assumptions of non
viscous fluid and translation symmetry. It can therefore be anticipated, or at
least it can be hoped, that these two functionals will be relevant in any
optimization principle\ for realistic large waves at the extreme position.

The major questionable point in the idea to generalize (\ref{minH_GM_real}) is
the role of the so-called `mass' functional $M$ that does not seem to
correspond to a physically well understood invariant functional in more
general situations. From its definition, $M\left(  S\right)  $ is seen to be
precisely the square root of the energy in the central frequency, $M\left(
S\right)  =S_{0}$. It will be shown in \cite{GroAnd05-VarExtWaves} that this
`mass' turns up as a special case, valid only at the extreme position, in a
variational principle that describes the complete SFB evolution as a relative
equilibrium according to general Hamiltonian theory. That variational
formulation depends on (the existence of) a higher order invariant functional,
and this seems to be related to the special properties of the completely
integrable NLS equation considered here. So the results presented here do not
unambiguously support the idea that the optimization principle for the extreme
signal can be expected to hold also in more realistic cases. Yet it is
tempting to look for such extremal formulations also in more realistic cases.

It may even be possible to investigate in a direct way the optimization
principle for the extreme signal from experiments in a well-controlled
laboratory environment. One possibility is to calculate the values of the
relevant functionals from the measured signal at the extreme position, and
investigate how close these values are near the optimal values from the
minimizing property. Another, less robust (and therefore maybe more
informative) method may be to use the fact that the extremizing property
reflects itself in a simple phase-plane representation of the signal, see
\cite{AKG}, so that the phase plane representation of the experimental extreme
signals could give an indirect indication. Further research in this direction
will be executed.

\textbf{Acknowledgement}

Regular discussions with Gert Klopman (UTwente) and Renee Huijsmans (MARIN)
have been helpful. Special thanks to Francesco Fedele for referring to
\cite{Fedele-manu04}\ and some useful discussions. Financial support from RUTI
(Ministry of Research and Technology, Indonesia) in the project `Wave motion
and Simulation', and support for visits to ITB of EvG and NK by EU-Jakarta
project SPF2004/079-057, and for visit of NK to 'Rogue Waves 04' by JMBurgers
Centre, is acknowledged. This work is also part of project STW 5374 `Extreme
Waves' from the technology division STW of NWO Netherlands.

\end{document}